\documentclass[pre,aps,twocolumn,superscriptaddress]{revtex4-2}

\usepackage[dvips]{graphicx}
\usepackage{amssymb,amsfonts,amsmath,gensymb}
\usepackage{color}
\usepackage[normalem]{ulem}
\usepackage{natbib}
\usepackage[hidelinks]{hyperref}
\usepackage{cleveref}
\usepackage{upgreek}

\newcommand{\ind}[1]{\ensuremath_{\mathrm{#1}}}

\newcommand{\C}{\ensuremath{ \mathcal{C}}}
\newcommand{\A}{\ensuremath{ \mathcal{A}}}

\newcommand{\sign}{\ensuremath{\mathrm{sign}}}

\renewcommand{\vec}[1]{\boldsymbol{#1}}
\newcommand{\Hp}{\ensuremath{\textbf{H}\ind{p}}}

\newcommand{\He}{\ensuremath{\textbf{H}\ind{ext}}}
\newcommand{\pa}{\ensuremath{\textbf{p}\ind{a}}}

\newcommand{\rv}{\ensuremath{\textbf{r}}}

\newcommand{\Vmag}{\ensuremath{V}_{\text{m}}}

\begin{document}

\title{Machine-learned topological transport of magnetic colloids}

\author{Sebastian Wohlrab} \affiliation{
  Physikalisches Institut, Universit{\"a}t Bayreuth, D-95440 Bayreuth, Germany}
\author{Aneena Rinu Perayil} \affiliation{
  Physikalisches Institut, Universit{\"a}t Bayreuth, D-95440 Bayreuth, Germany}
\author{Piotr Kuświk} \affiliation{
        Institute of Molecular Physics, Polish Academy of Sciences, 60-179 Poznań, Poland}
\author{Maciej Urbaniak} \affiliation{
        Institute of Molecular Physics, Polish Academy of Sciences, 60-179 Poznań, Poland}
\author{Feliks Stobiecki} \affiliation{
        Institute of Molecular Physics, Polish Academy of Sciences, 60-179 Poznań, Poland}
\author{Arne J. Vereijken} \affiliation{
        Institute of Physics, Center for Interdisciplinary Nanostructure Science and Technology (CINSaT), Universität Kassel, D-34132 Kassel, Germany}
\author{Arno Ehresmann} \affiliation{
        Institute of Physics, Center for Interdisciplinary Nanostructure Science and Technology (CINSaT), Universität Kassel, D-34132 Kassel, Germany}
        \affiliation{
        Artificial Intelligence Methods for Experiment Design (AIM-ED), Joint Lab of Helmholtzzentrum für Materialien und Energie, Berlin (HZB) and University of Kassel, Berlin, Germany
        }
\author{Thomas M.\ Fischer} \affiliation{
  Physikalisches Institut, Universit{\"a}t Bayreuth, D-95440 Bayreuth, Germany}
\author{Daniel de las Heras}
\email{delasheras.daniel@gmail.com}
\homepage{www.danieldelasheras.com}
\affiliation{Institute for Theoretical Physics, University of Tübingen, D-72076 Tübingen, Germany}
\affiliation{
        Physikalisches Institut, Universit{\"a}t Bayreuth, D-95440 Bayreuth, Germany}

\begin{abstract}
	By modulating the orientation of a uniform magnetic field it is possible to transport magnetic microparticles above magnetic patterns.
In periodic patterns, driving control loops that wind around special orientations, which are dictated by the symmetry of the pattern, induce transport by one unit cell.
The transport is topologically protected with trajectories controlled by the winding numbers of the control loop around the special orientations.
In non-periodic patterns, the special orientations are position dependent, precluding a global characterization of the transport using a small set of winding numbers.
We use here a reinforcement learning agent trained with Brownian dynamics simulation data that uses the local environment of the magnetic pattern to decide an action over the control loop.
The agent autonomously discovers topological transport protocols directly from the particle response without prior characterization of the magnetic potential landscape.
An agent trained solely on a complex non-periodic pattern directly generalizes to completely unseen periodic, quasicrystalline, and disordered substrates.
Moreover, the agent dynamically adjusts the control loop to transport particles across lattice defects and to counterbalance external drift forces that would cause static topological loops to fail.
By adding a pathfinding algorithm to the model, we additionally achieve simultaneous, independent transport of multiple identical colloids along different on demand trajectories.
We demonstrate that the control loops can be directly transferred to an experimental setup to successfully drive magnetic microparticles across a non-uniform magnetic thin film.
\end{abstract}

\date{\today}
\maketitle

\section{Introduction}

Reinforcement learning (RL)~\cite{s2018reinforcement} is a major branch of machine learning in which an agent autonomously learns the optimal behaviour (optimal policy) in a dynamical system.
The agent maps states of the system to actions and receives either positive or negative rewards based on its actions. 
Maximizing the cumulative reward, allows the agent to learn an optimal policy. 
The agent learns (maximizes the cumulative reward) through direct trial-and-error interactions with the system, rather than from a static training set as it is the case in other machine learning techniques.

In physics, reinforcement learning has found widespread applications across both classical and quantum systems, see recent reviews in Refs.~\cite{Yang2024,rev1,Banerjee2025}.
For instance, RL has emerged as a powerful tool to efficiently manipulating highly complex non-equilibrium systems.
RL agents can learn the optimal flocking strategy (velocity alignment) by using the local environment of the velocity field~\cite{PhysRevE.102.012601} illustrating the emergence of collective behaviour.
Smart microswimmers using a RL algorithm discover highly non-trivial navigation strategies to escape flow traps and maximize their altitude over time~\cite{PhysRevLett.118.158004}.
RL can discover spatiotemporal activity patterns that induce directed net transport in active systems and provide physically interpretable strategies for manipulating non-equilibrium many-body systems~\cite{PhysRevResearch.3.033291}.
Applying RL to active particles that communicate with each-other to maximize nutrient consumption, Grauer et al.~\cite{Grauer2024} identified three bio-inspired strategies: clustering, spreading, and adaptive behavioral switching.
Other recent works that make use of RL in physics include learning to train chemotactic navigation of microrobots~\cite{xiong2024enabling},
to control the collective behaviour of active systems subject to random noise~\cite{MuiosLandin2021}, for microswimmers to learn how to swim in nonuniform flow fields~\cite{Sankaewtong2023}, and 
for millimeter-sized magnetic robots driven by a time-varying magnetic field to learn how to navigate in complex environments~\cite{Behrens2022}.

A fundamentally different research direction in which RL has proven to be a versatile approach is topological protection, which refers to the robustness of physical phenomena against internal and external perturbations due to underlying topological invariants.
For instance, RL has enabled the optimization and control of topologically protected states such as magnetic skyrmions~\cite{Kim2026}, the discovery and characterization of topological states in quantum many-body systems~\cite{PhysRevB.96.195145}, and the design of novel topological materials~\cite{Xu2026}.

\begin{figure*}
        \centering
        \includegraphics[width=0.99\linewidth]{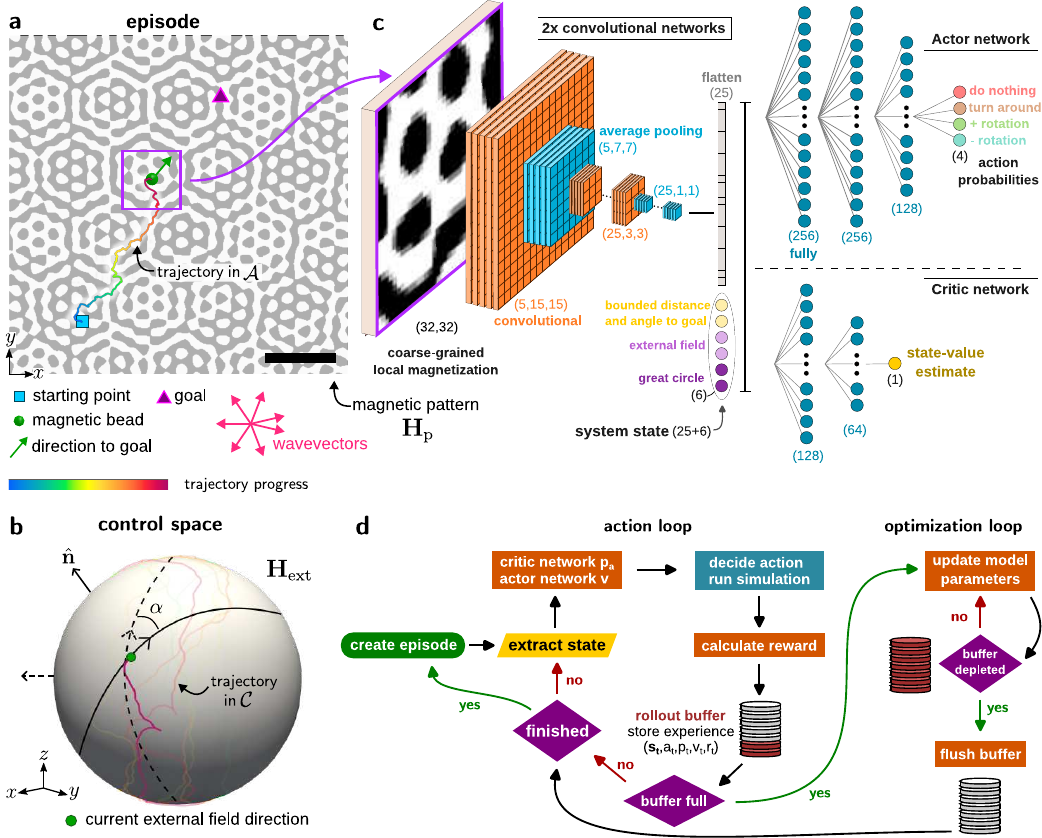}
        \caption{
	    {\bf System setup and reinforcement learning model.}
        (a) A section of the training pattern illustrating a training episode with its starting (blue square) and destination (purple triangle) points for the magnetic microparticle (green circle).
        Dark (white) regions denote positive (negative) magnetization along the direction normal to the pattern.
		The violet square denotes the local environment of the magnetization processed by the neural networks.
        The pink arrows show the directions of the wavevectors used to generate the pattern.
        The length of the scale bar (black) is $5a_0$.
		(b) Control space.
		The particle trajectory in action space (a) and that of the external field in control space (b) are colored using the same color code (color bar).
        To improve visualization, the trajectory in control space gradually fades, leaving only the final part of the modulation loop visible.
		(c) Neural network architectures.
		The sizes of the layers are indicated between parentheses.
		The first number in the convolutional layers indicates the number of channels.
		Each network, critic and actor, has their own convolutional network to preprocess the local environment of the magnetization.
		We use rectified linear units (ReLu) as the activation function of all nodes in all networks.
		(d) Workflow of the reinforcement learning training. 
        }%
 \label{fig1}
\end{figure*}

Here we make use of RL to control the topologically protected transport of magnetic microparticles.
In the basic setup, a magnetic particle is located above a thin magnetic pattern, Fig.~\ref{fig1}a, and driven by a time-dependent uniform external magnetic field, $\He(t)$.
The particle motion takes place on a plane parallel to the pattern that we refer to as action space $\cal A$. 
The magnetic pattern alternates between regions of positive and negative magnetization normal to the film, creating a static but space-dependent magnetic field $\Hp(\rv)$.
The pattern field is significantly weaker than the external field at the elevation at which motion occurs.

The magnetic particle is therefore subject to a space- and time-dependent magnetic potential, $\Vmag(\rv,t)\propto\He(t)\cdot\Hp(\rv)$, that results from the superposition between the external magnetic field and that of the pattern.
This simplified form of $\Vmag$ is due to the external field being much stronger than the pattern field at the elevation of the particle above the pattern.
For details on the calculation of $\Vmag$ we refer the reader to e.g. Refs.~\cite{Stuhlmller2023,Wohlrab2026}.
The energy landscape in the plane parallel to the magnetic film is complex and strongly dependent on the magnetization pattern.
In previous works~\cite{Loehr2016,Heras2016,Loehr2017}, we have studied the particle transport above periodic patterns such as checkerboard and hexagonal patterns.
The motion is driven by changes on the orientation of $\He$.
We define our control space $\cal C$ as the surface of a sphere, see Fig.~\ref{fig1}b.
Each point on $\cal C$ represents an orientation of $\He$. 
Particles are transported via the minima of the magnetic potential.
Changing the orientation of $\He$ in $\cal C$ alters the position of the minima of $\Vmag$ in $\cal A$ inducing particle motion.

For periodic patterns, the transport can be described with a topological invariant. 
There exist special orientations of $\He$ corresponding to critical points (bifurcation points), of the magnetic potential.
When a control loop in $\cal C$ winds around a bifurcation point, a particle is transported in action space (above the pattern) from one unit cell to a neighboring unit cell of the periodic pattern. 
While the detailed shape of the trajectories depends on the precise shape of the loops, the starting and ending unit cells are solely determined by the set of winding numbers around the special orientations in $\cal C$. 
The transport is topologically protected, and hence robust against internal and external perturbations, because it depends only on a topological invariant; the set of winding numbers of the modulation loop around the bifurcation points. 

However, a limitation of periodic patterns is that two particles located on different unit cells of the pattern perform parallel trajectories and hence their motion cannot be controlled independently. 
Non-periodic patterns can in principle be used to overcome such limitations and also admit a topological description of the transport.
So far, we have been able to control simultaneously and independently a collection of identical particles in simple non-uniform patterns~\cite{Stuhlmller2023}.
We create a grid with several periodic patterns (e.g. checkerboard patterns) in which each pattern of the grid is rotated with respect to the others. 
As a result, the control space of each pattern is also rotated, allowing for independent transport on each pattern of the grid.
Independent particle transport is still limited to particles that reside above two different subpatterns of the grid.

Overcoming such limitation requires using more complex non-uniform patterns, such as e.g. a quasicrystalline pattern.
However, understanding the transport and the topological properties on such pattern is a challenge because control space becomes space-dependent in an intricate way.
Essentially, each local environment of the magnetic pattern has its own set of bifurcation points in control space. 
We demonstrate here that reinforcement learning enables autonomous topological control in generic non-periodic patterns.
Using RL, we train an agent to transport particles above one non-periodic pattern.
We use a local-learning strategy in which the agent has access only to the local environment of the magnetic pattern at each time.
Due to this local-learning strategy, after training, the agent can successfully transport particles in any magnetic pattern without retraining.
Also, without retraining, the agent can respond to real-time changes, including the sudden appearance of defects in the magnetic pattern and the application of other external forces. 
Moreover, by adding a path-finding algorithm and retraining, the agent is able to perform simultaneous and independent multiparticle transport above generic patterns. 
The modulation loops are directly transferable to an experimental setup. 
The combination of reinforcement learning with topological physics shown here can find future applications on advanced lab-on-a-chip devices, automated targeted drug delivery platforms, and reconfigurable colloidal micro-machinery.

\section{System state and model architecture}

\noindent{\bf Training pattern.} To train the model, we construct a complex non-periodic pattern using a superposition of plane waves with prescribed wavevectors and amplitudes.
A representative section of the training pattern, covering approximately $\sim22\%$ of the total area is shown in Fig.~\ref{fig1}a.
We consider an infinitely thin pattern at $z=0$. The magnetization at position $\rv$ is given by
\begin{align}
	\mathbf{M}(\rv)=M(\rv_\perp)\delta(z)\hat{\mathbf e}_z,
\end{align}
with $\delta(\cdot)$ the Dirac distribution, $\rv_\perp=(x,y)$, and $\hat{\mathbf e}_z$ the unit vector along the $z$-direction (normal to the pattern). The sheet magnetization is
\begin{align}
	M(\rv_\perp) =m_{\rm p}\sign\left( \sum_{j=0}^{\lfloor n\rfloor} c_j^{(n)} \cos(\vec{q}_j^{(n)} \cdot \rv_\perp)\right).
\label{eq:mag}
\end{align}
Here, $m_{\rm p}$ is the sheet saturation magnetization (dimensions A), 
$\lfloor\cdot\rfloor$ indicates the floor operation, we select $n=2(1+\sqrt{5})$, and therefore $\lfloor n \rfloor = 6$.
The wave amplitudes are given by
\begin{align}
c_j^{(n)} =
\left\{
\begin{aligned}
1, &\qquad j \le \lfloor n \rfloor -2 \\
\frac{n-j}{2}, &\qquad j = \lfloor n \rfloor - 1\\
\frac{n-j+1}{2}, &\qquad j = \lfloor n \rfloor,
\end{aligned}
\right.
\end{align}
and the wavevectors are 
\begin{equation}
	\vec{q}_j^{(n)} = \frac{2\pi}{a_0}\left(\cos\phi_j^{(n)}, \sin\phi_j^{(n)}\right),
\end{equation}
with 
\begin{align}
\phi_j^{(n)} =
\begin{cases}
2\pi\left(1-\dfrac{j+1}{n}\right) + \phi_0,
& j \le \lfloor n \rfloor - 1 \\[6pt]
\phi_0,
& j=\lfloor n \rfloor,
\end{cases}
\end{align}
and
\begin{equation}
\phi_0 = \pi\left(1-\frac{\lfloor n\rfloor+1}{n}\right).
\end{equation}
Here, $a_0$ serves as our unit of length and establishes the characteristic scale of the magnetic domains (the side-length of the pattern region in Fig.~\ref{fig1}a is $\sim23a_0$).
The phase, $\phi_0$, simply aligns one wavevector with a Cartesian axis.
With the selected value of $n$, the pattern is generated with seven waves with non-equally spaced wavevectors, see Fig.~\ref{fig1}a.
The result is therefore a highly complex non-periodic magnetization pattern, especially since we chose $n$ to be a multiple of the golden ratio; the most irrational number.

We use proximal policy optimization (PPO)~\cite{PPO}, an on-policy reinforcement learning algorithm. 
Figure~\ref{fig1}c shows the neural network architecture of the model.
The algorithm uses the so-called actor-critic agents, where in addition to the neural network that controls the actions (in our case controlling the external magnetic field), there is a second network that evaluates the state of the system.
Both networks are trained individually using their own loss functions.
At each time, the input of both the critic and the actor networks is the instantaneous state of the system.
	
\noindent{\bf State of the system.} The instantaneous state of the system, $\mathbf{s}_t$, contains at each time $t$ the physical variables required to make an informed decision on the next value of the orientation of the external field.
These variables are the local environment of the magnetization, information about the relative position of the goal, and information about the current state in $\C$.

The local environment of the magnetization, see Fig.~\ref{fig1}c, is a square path of side-length $3.2a_0 $ centered around the position of the particle at time $t$.
In our implementation, the pattern is represented as a bitmap image ($1$ pixel represents $0.05a_0$) from which the local environment is extracted as a patch of side-length $32$ pixels. 
To reduce the size of the input, each pixel contains the average magnetization of the pattern in a square region of side-length $2$ pixels ($0.1a_0)$. 
After training, for handling patterns with higher or lower resolutions than the training pattern, either a pooling or a padding layer was used such that the input resolution remains unchanged. 
The size of the local environment is a trade-off.
It must be large enough (compare to $a_0$) to include information about adjacent stationary points of the potential, yet sufficiently small to both maintain computational efficiency and ensure that the policy generalizes to arbitrary patterns.

The local environment of the magnetization is processed by a convolutional network with two convolutional layers and two average pooling layers.
Each network in our PPO implementation (critic and actor) processes the local magnetization via its own convolutional network.
The output of the final average pooling layer is flattened and combined with the following six variables to form $\mathbf{s}_t$, the complete system state at time $t$: (1) the angle and (2) the (bounded) distance between the particle and the goal in $\A$,
(3) the polar and (4) the azimuthal angles of the external field in $\C$, and (5) the polar and (6) the azimuthal angles of the unit vector, $\hat{\mathbf{n}}$, in $\C$. This unit vector is normal to the great circle that defines the navigation direction of the trajectory in $\C$, see Fig.~\ref{fig1}b.
The angle between the particle and the goal is measured with repect to the $x$-axis. 
Rather than the Euclidean distance to the goal, $d_g$, we use a bounded distance $d_b=d_0\tanh(d_g/d_0)$ that saturates once $d_g\gg d_0$.
We set $d_0/a_0 =3$ and $d_0/a_0=5$ for single- and multi-particle transport, respectively. 
The bounded distance improves training stability because when the goal is far away, the exact distance is not relevant (the direction to the target is the governing variable).
Furthermore, the saturation enables direct generalization to cases where the distance to the goal is larger than those seen during training.
	
\noindent{\bf Model architecture.} The actor network (fully connected network with three hidden layers, Fig.~\ref{fig1}c) processes at each time the state of the system, $\mathbf{s}_t$, and outputs the discrete probability distribution, $\pa$, of four possible actions aimed at finding the next orientation of $\He$ in $\C$.
We use great-circle navigation to define the trajectory of the orientation of $\He$ in $\C$.
That is, the trajectory in $\C$ follows a great circle, Fig.~\ref{fig1}b, of varying orientation at constant angular speed ($0.05\,\text{rad}$ per step).
The possible actions, $a$, to update the trajectory are
\begin{align}
a &=
\begin{cases}
1, & \text{keep the current orientation of the great-circle} \\
2, & \text{turn around} \\
3, & \text{rotate the great circle by } +\alpha \\
4, & \text{rotate the great circle by } -\alpha.
\end{cases}
\end{align}
Larger values of $\alpha$ enable faster modulations adjustments, but at the expense of loosing resolution. We select here $\alpha=\pi/6$ as a compromise between fast adaptation and precise trajectory control in $\C$.

	The critic network (fully connected network with two layers, Fig.~\ref{fig1}c) processes the state of the system and outputs the state-value estimate, $v_t$, which represents an estimate of the expected cumulative reward from the current state of the system.
	That is, as explained below, the critic network estimates how good the current state of the system is. 

	\noindent{\bf Training.}
	The goal of the training is to adjust the $\sim10^5$ trainable parameters of the critic, the actor, and their corresponding convolutional networks such that a good set of actions (i.e.~those that move the particle to the desired destination) become more likely.    
	The workflow is illustrated in Fig.~\ref{fig1}d.
	The algorithm is split into the action and the optimization loop.
	In the action loop, we start by creating a new episode by generating random starting and destination points above the pattern.
	Both, the starting and destination points, must be selected within the so-called allowed regions of $\A$~\cite{Loehr2016,Heras2016}.
    For any point inside an allowed region, there exists an orientation of $\He$ that renders that point a minimum of the magnetic potential.
    
	It should be noted that we use only one pattern for training (a section of which is shown in Fig.~\ref{fig1}a).
	Although we explain here the process for one episode, to speed up the training, we use $50$ agents (instances of the actor network) that run episodes simultaneously.
        
	The state of the system, $\mathbf{s}_t$, is then computed and processed by the actor and the critic networks which output $\pa$ and $v_t$, respectively. 
	An action is then decided by sampling the discrete probability distribution $\pa$.
    Therefore, the taken decisions are not deterministic. 
    We designate $a_t$ and $p_{t}$ as the taken action and its corresponding probability, respectively.
	The orientation of $\He$ is updated according to the taken action, and the new particle position in $\A$ is calculated using Brownian dynamics (BD) simulations.
    In the experiments, the magnetic and viscous forces dominate and neither inertia nor Brownian motion play an important role.
    We, therefore, use the overdamped BD equation of motion in the limit of negligible thermal fluctuations to simulate the transport.
    For further details regarding the concrete calculation of the magnetic potential and BD implementation see Ref.~\cite{Wohlrab2026}.
    Every time $\He$ is updated, we run $100$ time steps, which guarantees that the particle finds the new minimum of the magnetic potential.
        
	Then, the instantaneous reward obtained by the taken action is calculated.
	We define a simple reward function, $r_t$, to measure how good the taken action was:
	\begin{align}
		r_t = 50\left(2\frac{\Delta_t}{r_{\text{tot}}}+h_t\right).\label{eq:reward}
	\end{align}
	Here, $\Delta_t$ is the difference between the distances to the goal before the action was taken and after the action was taken,
	$r_{\text{tot}}$ is the total distance between the starting and the destination points,
	and $h_t$ equals one if the particle reaches the goal (within a small tolerance) and zero otherwise.
	Therefore, Eq.~\eqref{eq:reward} rewards ($r_t>0$) actions that move the particle towards the goal and penalizes ($r_t<0$) actions that move particles away from the goal.
	Furthermore, it gives an extra reward if the goal is reached.
	The cumulative score is the sum of all rewards taken along the particle trajectory. 
	It follows from Eq.~\eqref{eq:reward} that the total possible cumulative score is $150$ which occurs if the particle reaches the goal. 
    Note that $r_t$ does not directly reward shorter trajectories in either action or control space.
    This allows the agent to better explore the configuration space and discover robust transport solutions.

	Next, we store the complete experience, $(\mathbf{s}_t,a_t,p_{t},v_t,r_t)$ in the so-called rollout buffer.
	The buffer accumulates $200$ experiences for each of the $50$ episodes that run simultaneously.
	We keep updating the trajectory generating new actions until the buffer is full.
	An episode ends if the particle reaches its destination point or after a number $a_{\text{max}}$ of actions taken. 
	When an episode ends, we start a new one. 

	Once the rollout buffer is full, we exit the action loop and enter the optimization loop, Fig.~\ref{fig1}d.
	There, we update the trainable parameters of all networks using the experiences collected in the rollout buffer to compute the loss functions.
	We use the Adam algorithm~\cite{Adam} for optimization with a learning rate of $3\times10^{-3}$.
	For the critic loss function we use
		\begin{align}
			L_\text{c} = \langle(\hat{A}_t + v_t - \pi(\{\phi\}|\mathbf{s}_t))^2\rangle,
		\end{align}
		where $\langle\cdot\rangle$ designates an average over the experiences in the batch,
		$\pi(\{\phi\}|s_t)$ is the output of the critic network for state $\mathbf{s}_t$ using the current set of trainable parameters $\{\phi\}$, and $\hat{A}_t$ is the generalized advantage estimator~\cite{schulman2015high} (we use the hyper-parameters $\gamma=0.9999, \Lambda=0.95$), which estimates how much better or worse an action $a_t$ performs in state $\mathbf{s}_t$ compared with the expected value of that state.
		The quantity $(\hat{A}_t + v_t)$ relates closely to what is often referred to as the ``return-to-go'', i.e. the sum of all rewards collected from this state onward.
		Roughly speaking, the critic network learns to predict the amount of rewards that can be gained from the current input state.
		
		For the actor network, we use the standard form of the actor loss in PPO with an additional entropy regularization term:
		\begin{align}
			L_\text{a} = L_\text{PPO-Clip} - 0.15 \langle H\rangle.
		\end{align} 
		Here, $L_\text{PPO-Clip}$ is the PPO-Clip loss function~\cite{PPO}, and $H$ is the Shannon entropy of the actor's action distribution.
		Minimizing the PPO-Clip loss function increases the probability of actions that produced good results and decreases the probability of bad actions~\cite{PPO}.
		The clipping (we use a clip hyper-parameter $\epsilon=0.2$) simply prevents the probability from changing too drastically in a single network update.
		The regularization term is minimum if the policy has high entropy and hence it encourages the policy to remain sufficiently random/exploratory.
		This aids in the exploration during training as well as makes the actor more reliable when solving unfamiliar problems after training.
	
	The PPO implementation was done in PyTorch~\cite{DBLP:journals/corr/abs-1912-01703}.
	The rollout buffer is used for $10$ batches with a batch size of $100$ randomly sampled experiences. 
	Afterwards, we flush the buffer and move back to the action loop.

    For single-particle transport, we train a single agent across three episode difficulty levels.
    While the actor network is shared across all levels, each level uses its own critic network.
    Using different critics makes the training more robust because our simple reward function, Eq.~\eqref{eq:reward}, is based on relative distances, and hence it does not distinguish whether the goal is close or not. 
	However, the optimal strategy might depend on the actual distance to the goal.
	Large distances to the goal might benefit from actions that induce large particle moves at the expenses of having less control over the position.
	On the contrary, short distances to the goal might benefit from actions that maximize control over the position at the expenses of being small particle moves.
	Once training is complete, only the actor network (which is the same for all types of episodes) is needed to generate the trajectory in control space.
	The three levels of difficulty are measured by the maximum allowed distance between the starting and the destination points, $r_{\text{max}}$:
	easy ($r_{\text{max}}/L=0.05$, $a_{\text{max}}=2\times10^3$),
	medium ($r_{\text{max}}/L=0.1$, $a_{\text{max}}=3\times10^3$),
	difficult ($r_{\text{max}}/L=0.6$, $a_{\text{max}}=10^4$). 
    Here, $L=50a_0$ is the side-length of the training pattern. 

	We use $10^3$ episodes of each difficulty level per each agent.
    Since we run $50$ agents simultaneously during traning, a total of $1.5\times10^5$ episodes are generated to obtain the optimal policy.
    The full training lasts less than one day on a RTX $4090$.
	The trained actor network successfully reaches the destination point in the training pattern with a probability larger than $0.98$.
    Because the actions taken by the agent are stochastic, separate runs sharing the same starting and destination points result in different control loops and hence distinct particle trajectories.

	\section{Results}
	As we show next, the trained model can be used to transport above general patterns single and multiple particles along complex trajectories.
        
        \subsection{Single particle transport}
	\begin{figure}
        \centering
        \includegraphics[width=0.99\linewidth]{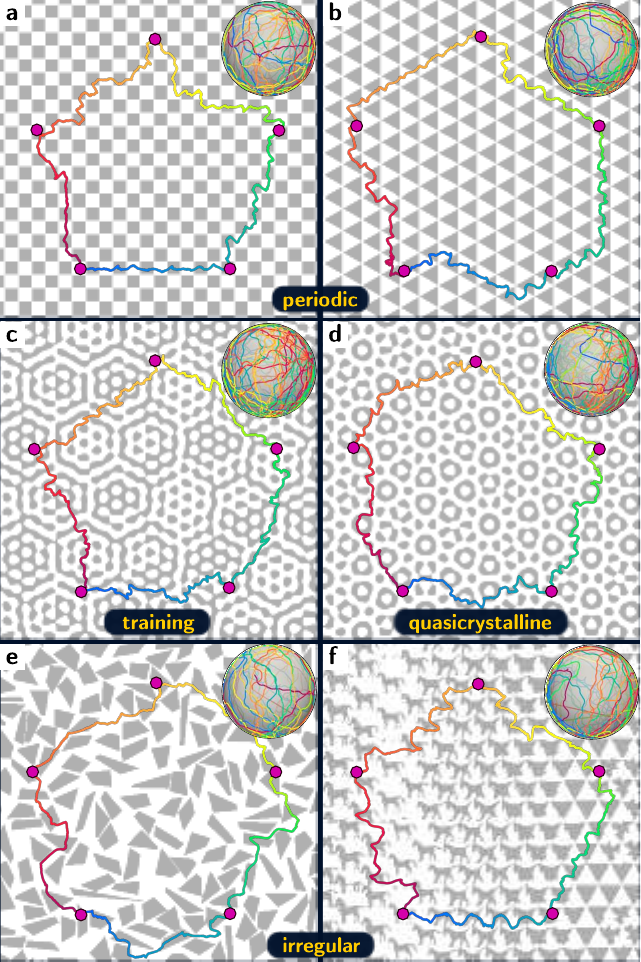}
        \caption{
		{\bf Single particle transport.}
		Particle trajectories and their corresponding trajectories in $\C$ for different magnetic patterns: periodic (a) four-fold and (b) three-fold symmetric patterns, (c) the training pattern, (d) a pattern with quasicrystalline order of the magnetic domains, (e) an irregular pattern generated with random polygons, and (f) an irregular pattern inspired by M. C. Escher's prints.
		The trajectories above the pattern and in $\C$ are colored according to their progress (see color bar in Fig.~\ref{fig1}a).
		The purple circles are the intermediate destination points that define the desired particle trajectory (a pentagon).
        A video of the particle trajectories and the corresponding loops in control space is available in the SM~\cite{SM}.
	         }%
	 \label{fig2}
	\end{figure}
We first evaluate the agent's ability to control particle trajectories above both the training pattern and alternative patterns.
Note that we do not retrain the agent.
Instead, we use the trained agent to transport particles above all patterns.
This is only possible due to the local-learning strategy in which the agent has access only to the local environment of the magnetization.
Illustrative results are shown in Fig.~\ref{fig2} for two periodic patterns with four-fold (a) and three-fold (b) symmetries, the original training pattern (c), a pattern with quasicrystalline order (d), an irregular pattern generated via random polygons (e), and another irregular pattern inspired by M.C. Escher's prints (f).
A video of the particle trajectories and the corresponding loops in control space is available in the Supplemental Material (SM)~\cite{SM}.

For each pattern, we design a closed trajectory defined by five intermediate destination points arranged as the vertices of a pentagon.
Each of these intermediate destinations corresponds to a minimum of the magnetic potential when the uniform external field is oriented parallel to the pattern normal.
Therefore, the exact location of each intermediate point depends on the local magnetization environment and varies slightly across the different patterns.

In all cases, the agent successfully transports the particle along the desired trajectories.
The corresponding loops in control space are notably complex, see Fig.~\ref{fig2}.
In particular, the loop in control space for the training pattern appears more intricate than those of the other patterns (note that the loop covers almost the complete control space).
This increased transport complexity on the training pattern was to be expected compared to periodic patterns but it is a nontrivial result for the other patterns.
Transport within the training pattern seems to be more challenging, which together with the local-learning approach, explains why an agent trained exclusively on this specific pattern can generalize and successfully transport particles across any other pattern.
The training pattern contains a sufficiently rich variation of local magnetization environments to enable robust generalization to other patterns.

\begin{figure}
        \centering
        \includegraphics[width=0.91\linewidth]{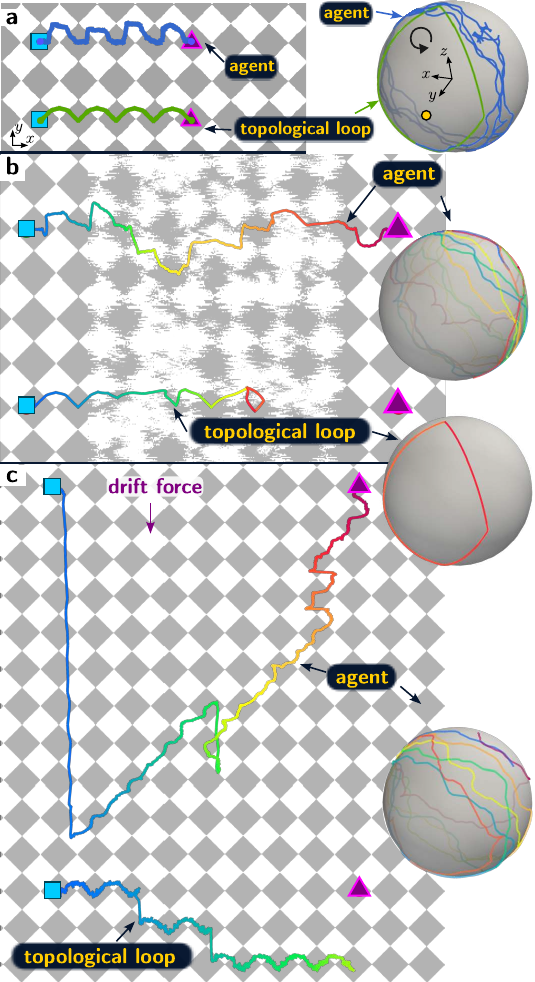}
        \caption{
		{\bf Impeded transport.}
		Particle trajectories and corresponding loops in $\C$ for (a) a periodic four-fold symmetric (checkerboard) pattern,
        (b) a checkerboard pattern with a distorted central region, and
        (b) a checkerboard pattern in presence of Brownian motion and a drift force, the direction of which is indicated by the violet arrow. 
		The trajectories above the pattern and in $\C$ in (b) and (c) are colored according to their progress (see color bar in Fig.~\ref{fig1}a).
	    Two particle trajectories are shown in each panel: the agent and the topological trajectory, as indicated.
        The yellow circle in $\C$ (a) indicates the position of the bifurcation point along $+\hat{\mathbf e}_y$.
		The blue squares and the pink triangles are the starting and destination points of the trajectories, respectively. 
		A video of the particle trajectories in panels (b) and (c) with the corresponding loops in control space is available in the SM~\cite{SM}.
		  }%
	 \label{fig3}
	\end{figure}

The agent learns which loops are topologically non-trivial, inferring the topology directly from the particle response rather than from the characterization of the magnetic potential via action space space and control space. 
To illustrate this, we compare the modulation loops generated by the agent with conventional topological loops for a simple task: transporting a particle along the direction of a lattice vector in a four-fold symmetric pattern, see Fig.~\ref{fig3}a.
In this example, to induce transport along $+\hat{\mathbf{e}}_x$, the driving loop in $\C$ must wind counterclockwise around $\hat{\mathbf{e}}_y$~\cite{Heras2016,Loehr2017,Wohlrab2026}.
Transport is more robust when the loop remains far from the bifurcation points in $\C$, which for four-fold symmetric patterns are four equidistant points on the equator of $\C$ aligned with the lattice vectors of the pattern.
The topological loop (green) shown in Fig.~\ref{fig3}a is an example of a robust loop.

Because the agent's policy is stochastic, we evaluated 5000 independent runs, all of which succeeded, and identified a particularly robust path as the shortest trajectory in action space.
The modulation loop discovered by the agent (blue) also winds around the same axis in $\C$ while staying away from the bifurcation points.
The path length in action space can be a good indication of the robustness of the transport.
The closer a modulation loop approaches a bifurcation point in $\C$, the faster the potential minima shift in action space.
Hence, short paths in $\A$ will in general correspond to loops in $\C$ that stay as far away as possible from bifurcation points.

The ability to transport particles above arbitrary geometries gives the agent an immediate advantage over the use of predefined, static, topological loops.
Particle transport succeeds even when the pattern contains imperfections, which in the experimental setup can arise from fabrication defects or pattern degradation.
An example of this capability is shown in Fig.~\ref{fig3}b, where we compare the performance of the agent (without retraining) against topological loops for transporting particles across a checkerboard pattern with a distorted central region.
The task is simply to translate the particles a few unit cells along the direction of a lattice vector.
The specific topological loop used here is highly robust by design, as it avoids passing near the bifurcation points of the checkerboard pattern in control space~\cite{Heras2016,Loehr2017}.
However, despite this robustness, the particle driven by the topological loop fails to reach the destination and instead gets trapped within the distorted region of the pattern.
In contrast, the agent adjusts the loop in control space to the changing environments of the local magnetization encountered along the path and it successfully transports the particle to the goal.

The agent is also capable of correcting the trajectory when new external forces act on the particle, as illustrated in Fig.~\ref{fig3}c.
In this case, in addition to the magnetic forces, we apply both a constant drift force (fixed magnitude and direction) and a random force that varies at each time step.
Experimentally, such a drift force can arise e.g.~from gravity if the pattern is not perfectly horizontal.
The random force mimics Brownian motion, which is typically negligible for the beads used in our experiments (particle diameter of several micrometers) but becomes significant as the particle scale decreases.

The checkerboard pattern is a difficult test for the agent due to the presence of flat channels in the magnetic potential~\cite{Heras2016,MirzaeeKakhki2020b,Stuhlmller2022}. 
When the external field points along a bifurcation point of the four-fold symmetric pattern in control space, the magnetic potential becomes flat along the direction of a reciprocal lattice vector in action space.
Under these conditions, even a weak drift force can significantly alter the particle's trajectory, as illustrated in Fig.~\ref{fig3}c.
Shortly after the trajectory begins, the particle controlled by the agent enters one of these flat potential channels, causing a strong deviation of several unit cells perpendicular to the intended path.
Despite this initial detour, the agent successfully regains control of the trajectory and guides the particle to the goal.
In contrast, the topological loop fails.
Even though the loop is robust and avoids the bifurcation points in control space, the drift force is strong enough to induce unintended particle transport perpendicular to the direction of the topological transport.

\subsection{Multiparticle transport}

\begin{figure*}
\centering
\includegraphics[width=0.99\linewidth]{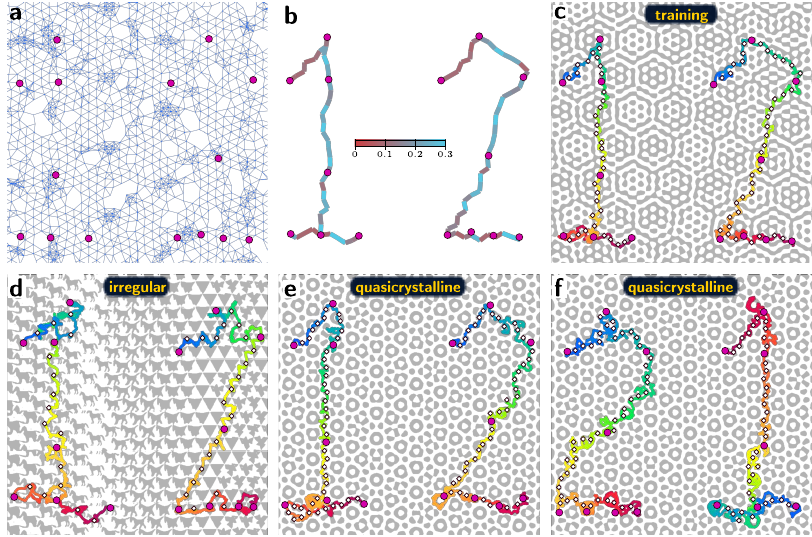}
\caption{
    {\bf Simultaneous transport of two particles.}
	(a) Minima of the magnetic potential (blue dots) when the external potential is oriented parallel to the pattern normal, $\He\parallel\hat{\mathbf e}_z$. Neighboring minima are connected by line segments.
    Desired particle trajectories are defined via intermediate destination points (magenta circles).
	(b) Optimal particle paths joining consecutive destination points according to the pathfinding algorithm and color-coded according to the estimated success rate by the ranking network (see color bar).
    Note that in some segments of the paths only one particle moves while the other stays in the same position.
    Hence, there is not a one-to-one color match between the segments of both particle paths in (b).
	Two particle trajectories tracing the numbers $1$ (left) and $2$ (right) above (c) the training pattern, (d) an irregular pattern, and (e) a quasicrystalline pattern.
    (f) Two particle trajectories tracing the numbers 2 (left) and 1 (right) on the quasicrystalline pattern.
	The trajectories are colored according to the progress of the control loop (see color bar in Fig.~\ref{fig1}a).
    Note that in panel (f) the two particles trace the trajectories in opposite directions.
    The small white circles in panels (c) to (f) are the optimal set of local minima of the potential that form the trajectories as selected by the path-finding algorithm.
    A video of the particle trajectories and the corresponding loops in control space is available in the SM~\cite{SM}.
    }%
\label{fig4}
\end{figure*}

Particles belonging to different topological classes can be transported simultaneously and independently (i.e., along distinct trajectories) using a single control loop because the positions of their bifurcation points in control space differ.
Examples include the simultaneous transport of diamagnetic and paramagnetic particles~\cite{Loehr2016}, bipeds (chains of beads) of different lengths~\cite{MirzaeeKakhki2020}, and isotropic particles of different sizes~\cite{Wohlrab2026}.
In contrast, identical particles belong to the same topological class.
Therefore, to transport them simultaneously and independently, the particles must be located above different local magnetization environments.
Otherwise, the relative transport induced by the external field would be the same for identical particles.

In Ref.~\cite{Stuhlmller2023}, we demonstrated that controlling several identical particles simultaneously is feasible by using a pattern composed of a grid of multiple checkerboard patterns, where each subpattern is rotated with respect to the others.
Because of these rotations, the bifurcation points in control space are also rotated, allowing particles residing above different checkerboard domains to be controlled individually via a single control loop.
However, a drawback of this method is that two particles situated above the same checkerboard domain always follow parallel trajectories.
Furthermore, initializing the particles above the grid is non-trivial, requiring, for instance, the use of secondary patterns that feature a topological defect in the magnetization acting as a particle attractor.
From these defect positions, the particles must then be transported to their respective checkerboard domains~\cite{Stuhlmller2023}.

Here, we demonstrate that the agent can simultaneously and independently move identical particles above complex patterns.
This approach offers advantages over the use of static topological loops on grid patterns.
First, it eliminates the need to initialize particles using complex secondary patterns.
Second, instead of a grid of distinct domains, the system uses a single pattern with a spatially varying, complex magnetization.
Hence, simultaneous trajectories can be controlled across the entire pattern rather than being restricted to separate domains of the grid.

	\begin{figure}
        \centering
        \includegraphics[width=1.00\linewidth]{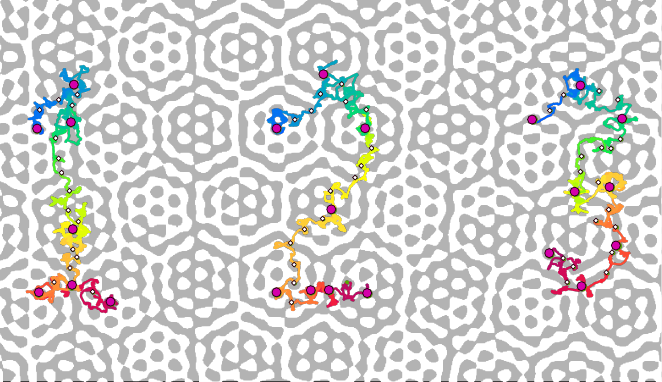}
        \caption{
		{\bf Multiparticle transport.}
		Simultaneous transport of three particles on top of the training pattern.
                The big purple circles are the intermediate destination points that define the three desired particle trajectories (a number from $1$ to $3$ for each particle).
		The small white circles are the intermediate steps between two consecutive destination points as decided by the path finding algorithm.
		Each particle traces a number from $1$ to $3$. 
		The trajectories are colored according to the progress of the control loop (see color bar in Fig.~\ref{fig1}a).
		A video of the particle trajectories and the corresponding loop in control space is available in the SM~\cite{SM}.
	         }%
		 \label{fig5}
	\end{figure}

We begin training an agent capable of transporting two identical particles simultaneously and independently from two initial positions to two desired goals.
To achieve this, the RL model previously described for single-particle transport, Fig.~\ref{fig1}, must be modified.
First, the convolutional networks of the actor and critic networks process the local magnetization environments of both particles.
The flattening layer then integrates the features from both local environments with the current external field and the current angles and distances to the respective goals.
In principle, this minimal modification is sufficient to train a model capable of multiparticle transport.
However, for complex trajectories, the success rate for both particles reaching their destinations is relatively low.
For this reason, we introduce additional modifications to our method: a ranking network together with a pathfinding algorithm.

Figure~\ref{fig4} illustrates the simultaneous transport of two particles.
We initially find the minima of the magnetic potential when the external field is oriented parallel to the pattern normal (blue dots in Fig.~\ref{fig4}a). 
Occasionally, a minimum is relatively flat and it is therefore not properly detected by our numerical algorithm (note the white/empty regions in Fig.~\ref{fig4}a). 
Nevertheless, the density of minima when $\He\parallel\hat{\mathbf e}_z$ is sufficiently high that even if not all minima are detected, the particle trajectories can still be defined with high precision.

Training then continues with elementary episodes where either both particles move to neighboring minima, or one particle stays in place while the other moves to an adjacent minimum.
The ranking network is trained alongside the actor and critic networks, aiming to predict the success rate of the episode.
It is a fully connected network that takes the starting state of an episode as input and outputs an estimated success rate.
Note that the state includes the distance and direction to the goal.
The loss function is simply the mean square error to the actual success rate of the problem.
We also increased the capacity of the rollout buffer to accumulate $1000$ experiences since long-term strategy is more relevant to complete a multiparticle transport problem.

After training, the ranking network is used to assist the pathfinding algorithm in order to efficiently navigate transport problems where the starting and ending positions are non-neighboring minima in the potential.
As before, we define the desired particle trajectories via intermediate destination points, see Figs.~\ref{fig4}a.
We then decompose the trajectories between any two consecutive destination points into smaller segments, mimicking the problems generated during training in their complexity.
That is, transporting each particle at most to a neighboring minimum.

A pathfinding algorithm selects the optimal trajectory between destination points based on the success rate estimated by the ranking network.
We employ a standard $\rm{A}^*$-pathfinding algorithm~\cite{hart1968formal} on a graph, where each node represents the position of both particles, and edge weights are computed from the output of the already trained ranking network. 
Weights represent the difficulty of moving from one node to another, with high weights indicating that moving along the corresponding  edge is difficult.
We set each edge weight to the inverse of the estimated success rate (the output of the ranking network). 
An alternative and faster option to transport on sufficiently simple patterns is to use constant edge weight for any estimated success rate above a given threshold, essentially only disallowing edges that represent especially hard problems.
In either case, it is important that node connections where the ranking network predicts a near zero success rate are forbidden.
To reduce computation time, the navigation graph only includes the minima within an ellipse around the starting and ending positions of each particle.
$\rm{A}^*$-pathfinding uses an heuristic function to estimate the cost from a given point to the goal.
Here, we use for the $\rm{A}^*$-heuristic function the sum of position-end distances.
The pathfinding implementation was created with the python network analysis package Networkx~\cite{hagberg2020networkx}.

Figure~\ref{fig4}a shows the collection of intermediate destination points (magenta circles) for both particles.
The optimal path, which is determined between consecutive destination points by the pathfinding algorithm, is shown in Fig.~\ref{fig4}b.
This optimal path passes through a series of local potential minima and minimizes the cumulative path difficulty as evaluated by the pathfinding algorithm.
The optimal path in Fig.~\ref{fig4}b is color-coded according to the estimated success rate for each segment.
As expected, path segments where the two local environments and the two instantaneous directions of motion are similar for both particles exhibit high estimated success rates.
If either of these conditions is not satisfied, the estimated success rate can vary substantially. 
However, the agent consistently finds a control loop that induces the desired transport.

Figure~\ref{fig4}c shows the actual simultaneous trajectories of two particles above the training pattern.
Both particles successfully reach all intermediate destination points along the desired trajectories, tracing the shapes of the numbers $1$ (left) and $2$ (right).
Notably, we do not retrain the convolutional network that processes the local magnetization environments.
Instead, we directly use the network already trained for single-particle transport. 
Moreover, the multitransport agent is again able to simultaneously transport particles above unseen patterns without any retraining.
As an example, we deploy the agent to move two particles along the same trajectories (tracing the numbers $1$ and $2$) but above an irregular pattern, Fig.~\ref{fig4}e, and a quasicrystalline pattern, Fig.~\ref{fig4}d.
The exact position of the intermediate destination points (violet circles in the figure) vary slightly from pattern to pattern because they are constrained to be local minima of the magnetic potential.
Nevertheless, the resulting trajectories are quite similar across the three patterns, demonstrating the robust capability of the agent to transport particles above general patterns.
Furthermore, the agent provides full control over the individual trajectories, as illustrated in Fig.~4f.
Here, we swap the paths: the left particle traces the number $2$ while the right particle traces the number $1$.
Additionally, the tracing directions are reversed, with number $2$ drawn from top to bottom and number $1$ from bottom to top.
Driving the particles in opposing directions for the majority of the path is a significantly harder task.
While this requires a control loop twice as long as in the previous case, the agent reliably guides both particles through their targets.

In principle, it is possible to transport several particles simultaneously, but the complexity increases significantly with the number of particles. 
As a proof of concept, we present in Fig.~\ref{fig5} the simultaneous transport of three particles tracing the numberVelascos $1$, $2$, and $3$ on the training pattern.
To transport three particles simultaneously, we extended and retrained the model to incorporate information from all three particles, in direct analogy to the previous modification from single- to two-particle transport.

\subsection{Experiments}
To provide an experimental proof of concept and validate our RL framework, we demonstrate single-particle transport across a non-uniform magnetic pattern.
The pattern is closely related to the training pattern because it was generated using $n=6.5$ in Eq.~\eqref{eq:mag}.
The magnetic film is a Ti(4 nm)/Au(60 nm)/[Co(0.7 nm)/Au(1 nm)]$_5$ multilayer that was magnetron sputtered on a naturally oxidized Si(100) substrate.
Magnetic patterning is achieved by keV $\mathrm{He}^+$-ion bombardment through a lithographical mask~\cite{Chappert1998,Mougin2001,Kuswik2011}.
The pattern is coated with a protective polymer layer (thickness $1.0\,\mu\mathrm{m}$) that also acts as a spacer.
Dynabeads (diameter $2.8\,\mu\mathrm{m}$) with a polystyrene shell and a core consisting of polymer microspheres containing superparamagnetic nanoparticles are suspended in deionized water and sediment under gravity. 
The particles, which effectively behave like paramagnetic particles, reach an equilibrium elevation at a distance on the order of the Debye length (tens of nanometers) above the polymer layer.
The equilibrium elevation is governed by the balance between gravitational, magnetic and DLVO~\cite{Wirix2005} forces.
The uniform external magnetic field, $\He$, is generated by a triaxial coil assembly interfaced with a control computer.
The field magnitude is approximately $4\times10^3\,\mathrm{A}/\mathrm{m}$.
Both the colloidal particles and the underlying magnetic pattern are imaged simultaneously via reflection microscopy.
Further details regarding the experimental setup are available in e.g.~Ref.~\cite{Loehr2017}.

\begin{figure}
	\centering
	\includegraphics[width=0.95\linewidth]{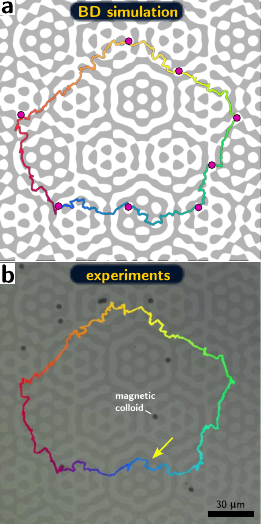}
	\caption{
        {\bf Experimental proof of concept.}
        Trajectories of a colloidal particle obtained from (a) Brownian dynamics simulations and (b) experiments.
        Purple circles mark the intermediate destination points defining the desired trajectory (a pentagon).
        The trajectory color indicates temporal progress (see color bar in Fig.~\ref{fig1}a).
        The arrow indicates the region exhibiting the largest difference between simulation and experiment.
        The total time to complete the trajectory in the experiments was approximately $700\,\mathrm{s}$.
        The particle diameter is $2.8\,\mu\mathrm{m}$.
        The scale bar spans approximately $30\,\mu\mathrm{m}$.
        A video showing both particle trajectories and the corresponding loop in control space is provided in the SM~\cite{SM}.
	}%
	\label{fig6}
\end{figure}

In principle, the RL agent could control the experimental setup directly, dynamically adjusting the orientation of $\He$ in response to real-time particle tracking instead of training within BD simulations.
In our current setup, however, the agent is trained with simulation and subsequently the control loops are transferred to the experiment.
This mode of operation is required by hardware constraints since our wave generators (three independent units to control the Cartesian components of $\He$) require the full modulation loop in $\C$ to be preloaded into memory before execution.
Hence, the orientation of $\He$ cannot be altered on the fly, which precludes a direct control of the experimental setup by the RL algorithm.
Additionally, the control computer interfacing with the wave generator hardware lacks the computational throughput required for machine learning and computer-vision tracking.

We first compare the live microscopy image with the theoretical magnetization pattern to localize the particle's coordinates of a target colloidal particle.
This configuration is then mapped into our BD simulation environment, where the RL agent identifies the control loop required to (i) drive the particle to the desired starting point and (ii) execute the target trajectory.
An illustrative proof-of-concept realization is presented in Fig.~\ref{fig6}, which directly compares the simulated (a) and experimental (b) trajectories along a pentagonal path. 
The two paths show remarkable agreement, with transport proceeding across the same sequence of potential minima in the magnetic landscape.
The only discernible discrepancy emerges in a localized region (highlighted by the yellow arrow in Fig.~\ref{fig6}b), where the simulated and experimental particles temporarily occupy different adjacent local minima.
This divergence, however, does not compromise the global transport, and the particle reliably returns to the target trajectory.

Note that the experimental trajectory shown in Fig.~\ref{fig6}b represents a piecewise concatenation of four separate experimental segments.
In the video included in the SM~\cite{SM}, each transition between segments is explicitly indicated and one can appreciate how only the target particle follows the desired trajectory due to the magnetic pattern being non-periodic.
Whenever a segment terminates, the external magnetic field needs to be turned off, temporarily flattening the energy landscape.
To resume transport, we identify a colloidal particle situated nearby and use the RL agent to generate a control loop that guides it to the terminal point of the previous segment before continuing with the target trajectory.
Completing the entire trajectory within a single experiment is currently not possible by two technical constraints.
First, the length of the complete control loop exceeds the buffer capacity of the wave generators.
Second, real-time changes to the control loop are not possible.
Consequently, experimental disruptions, such as unexpected particle pinning or doublet formation induced by dipolar interactions, cannot be corrected dynamically on the fly.

Our current technical constraints are by no means fundamental.
Updating the wave generators and integrating real-time particle tracking will enable direct coupling between the RL algorithm and the experimental setup.
Nevertheless, it is remarkable that our RL policy trained purely on Brownian dynamics simulations achieves such precise control over a complex physical experiment without retraining.
This demonstrates the robust simulation-to-experiment transferability of the learned magnetic driving protocols, which is possible due to the topologically protected nature of the transport.

\section{conclusions}
We have demonstrated that reinforcement learning provides a powerful tool for controlling the topologically protected transport of colloidal particles above complex magnetic patterns.
An agent makes use of the local environment of the magnetization to decide an action over the control loop that drives the motion.
Trained exclusively on a single non-periodic pattern generated from incommensurate wavevectors, the agent exhibits remarkable zero-shot generalization.
That is, without any retraining or parameter tuning, the agent successfully drives colloids across general patterns including periodic lattices with different rotational symmetries, quasicrystalline arrangements, and fully disordered substrates.
Instead of a global characterization of the topological properties of the transport for each pattern, the agent (using a local-learning approach) is able to infer the topology directly from the particle response, generalizing topological transport to aperiodic (disordered) systems.
Additionally, the modulation provided by the agent induces more robust transport compared to that of static topological loops.
The agent dynamically adjusts the control loop to avoid localized structural defects of the pattern and to counterbalance other external forces that otherwise could make static topological loops fail.

Furthermore, we can also simultaneously and independently control identical particles in arbitrary patterns.
By coupling the RL policy with a ranking network and a pathfinding algorithm, the system decomposes complex global trajectories into sequences of elementary transitions between consecutive potential minima, enabling simultaneous multi-particle transport along independent paths.
This approach is conceptually similar to hierarchical planning in robotics~\cite{yamamoto2018,Eppe2019,woehlke2021}. 
A low-level RL agent governs local transport between adjacent potential minima, while a high-level pathfinding algorithm plans the global trajectory.
Crucially, the high-level planner uses knowledge of the low-level dynamics to find the optimal route.
In contrast to other techniques that require targeted forces, our approach uses a single, uniform magnetic field to simultaneously drive identical particles along distinct, independent trajectories.

We have tested the viability of our methodology experimentally by directly transferring the policy learned with Brownian dynamics simulations to a physical setup in which we have achieved highly accurate transport of magnetic colloids on thin magnetic films.

For future work we plan to connect the simulation-trained RL framework directly to real-time particle tracking in the experiments, which will enable autonomous colloidal manipulation.
The agent will then be able to respond to real-time changes.
If something unexpected occurs during the transport, the agent could reroute the loop to maintain transport.
This real-time adjustments can be important if fabrication defects or degradation of either the magnetic pattern or the polymer coating are present, and also if the experiment is externally perturbed.
For example, an inclination of the pattern can add a drift force (gravitational field) on the particles. 
It is possible to use the experimental local environment of the magnetization because the pattern is visible under the microscope.
However, we have seen that the agent compensates for external forces absent during training.
We, therefore, expect that using the idelized theoretical magnetization patterns will be sufficient to control the transport in experiments.

Beyond the single- and multi-particle transport considered here, we are confident that a similar framework can be deployed to independently and simultaneously control the motion of microparticles of different sizes~\cite{Wohlrab2026}, different magnetic properties~\cite{Loehr2016}, and different shapes~\cite{MirzaeeKakhki2020} on arbitrary magnetic patterns, as well as to control the transport of active magnetic particles~\cite{Jin2021} including the guidance of magnetotactic bacteria on magnetic substrates~\cite{Loehr2016bacteria}. 
Our topological description of the transport should be directly transferable to ferroelectric domain patterns driven by modulated electric fields.
Nanoscale ferroelectric domains~\cite{Catalan2012,Yadav2016} could enable topological transport of polar nanoparticles.
The generalization capabilities of the RL agent could be crucial in overcoming challenges inherent to the nanoscale.
For example, beyond the thermal fluctuations that directly perturb particle trajectories, temperature can also induce structural changes to the pattern itlsef~\cite{Ortiz2022}.

\section*{acknowledgments}
D. d. l. H. acknowledges support through the Heisenberg program of the Deutsche Forschungsgemeinschaft (DFG, German Research Foundation) under Projects No. 550390029.
D. d. l. H.  and T. M. F. acknowledge funding by DFG under Project
No. 531559581 and No. 583739578.
A. E. acknowledge funding by the DFG under Projects No. 514858524, No. 566304473, No. 433501699, No. 361379292, and No. 361396165.
P. K., M. U., and F. S. acknowledge financial support from the National Science Center Poland through OPUS funding (Grant No. 2020/39/B/ST5/01915).

\section*{Author contributions}
S. W. implemented the BD simulation framework, developed and trained the RL model, and designed the training pattern.
A. R. P. performed the experiments and tracked the particle trajectories.
P. K., F. S., and M. U. produced the magnetic film, optimized the magnetic properties and produced the layered system.
A. J. V. and A. E. performed the fabrication of the micromagnetic domain patterns within the magnetic thin film.
S. W. and D. d. l. H. conceived the RL model and simulation framework, and A. E. and T. M. F. conceived the experimental setup.
S. W. and D. d. l. H. wrote the manuscript with input from all the other authors.

\end{document}